# Sustainability of Data Center Digital Twins with Reinforcement Learning

**Soumyendu Sarkar**[*,†], **Avisek Naug**[†], **Antonio Guillen**[†], **Ricardo Luna**[†], **Vineet Gundecha**[†],
**Ashwin Ramesh Babu, Sajad Mousavi**

Hewlett Packard Enterprise

{soumyendu.sarkar, avisek.naug, antonio.guillen, rluna, vineet.gundecha, sajad.mousavi, ashwin.ramesh-babu}@hpe.com

**Abstract**

The rapid growth of machine learning (ML) has led to an increased demand for computational power, resulting in larger data centers (DCs) and higher energy consumption. To address this issue and reduce carbon emissions, intelligent design and control of DC components such as IT servers, cabinets, HVAC cooling, flexible load shifting, and battery energy storage are essential. However, the complexity of designing and controlling them in tandem presents a significant challenge. While some individual components like CFD-based design and Reinforcement Learning (RL) based HVAC control have been researched, there's a gap in the holistic design and optimization covering all elements simultaneously. To tackle this, we've developed DCRL-Green, a multi-agent RL environment that empowers the ML community to design data centers and research, develop, and refine RL controllers for carbon footprint reduction in DCs. It is a flexible, modular, scalable, and configurable platform that can handle large High Performance Computing (HPC) clusters. Furthermore, in its default setup, DCRL-Green provides a benchmark for evaluating single as well as multi-agent RL algorithms. It easily allows users to subclass the default implementations and design their own control approaches, encouraging community development for sustainable data centers. Open Source Link: https://github.com/HewlettPackard/dc-rl

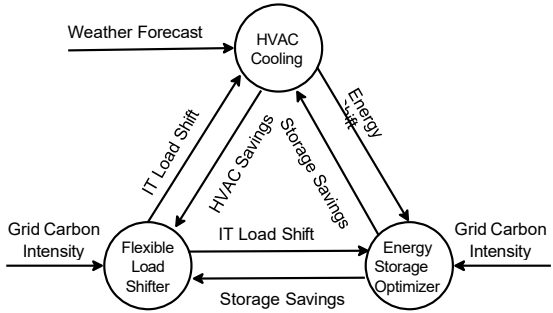

Figure 1: Internal and External System Dependencies

---

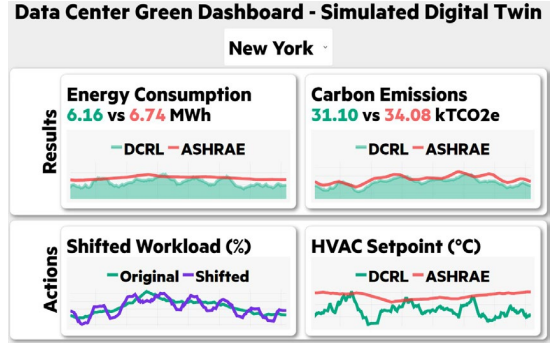

Figure 2: Sustainability Dashboard for Data Centers

## Introduction

In order to reduce carbon emissions, data centers have to be efficient with respect to design and control. Recent advances in Reinforcement Learning (RL) (3; 23; 18; 16; 19; 20; 17; 24; 10; 11; 14; 21; 13; 12; 26) have led to the creation of key RL environments to optimize energy and resource allocation in building and data center management (27; 25; 6; 2; 1; 9). While they focus on individual approaches like flexible load shifting, HVAC setpoint optimization and battery scheduling, none of them jointly optimize the behavior of these individual approaches in real-time due to the difficulty in modeling the complex interdependencies (Figure 1). Our work achieves this through the DCRL-Green framework (Figure 2).

## DCRL-Green

DCRL-Green is a tool for evaluating designing carbon-efficient data centers and single as well as multi-agent reinforcement learning strategies aimed at reducing data center carbon footprints. Built on the OpenAI Gym, this comprehensive framework highlights:

**Simulation Framework** (Data Center Digital Twin in Figure 3): DCRL models multizone zone IT Rooms, where the user can provide custom IT Cabinet geometry, server power specifications, and an HVAC system comprising chillers, pumps, and cooling towers (7; 8). It also has default as well as easily customizable models for grid carbon intensity aware Load Shifting Model and auxiliary Battery supply. The models seamlessly integrate with external software like EnergyPlus or directly in Python.

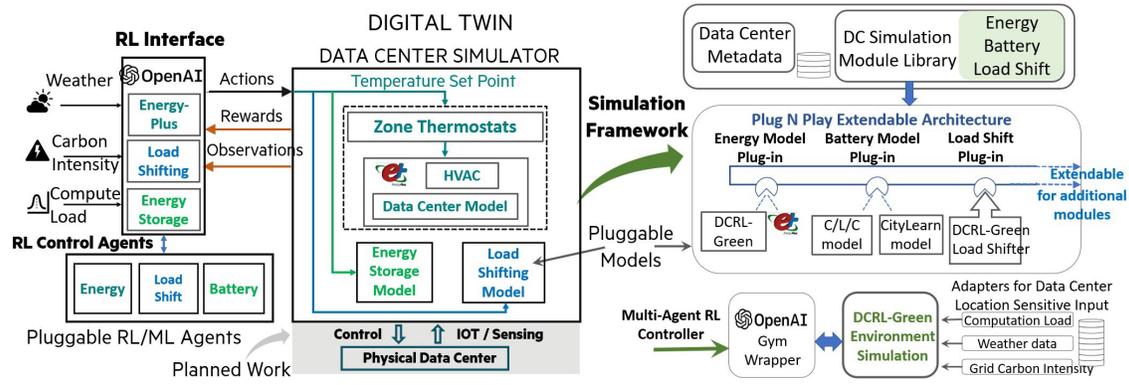

Figure 3: DCRL-Green Data Center Digital Twin Configurable Modeling.

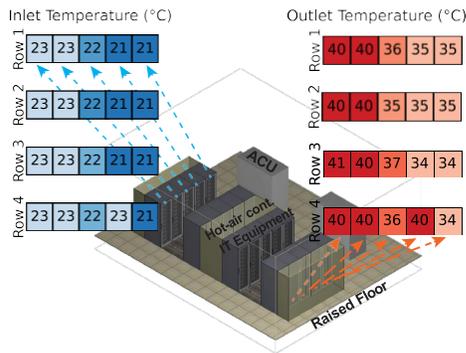

Figure 4: Temperature Distribution generated from a given Configuration of Data Center

**Configurability** (Plug N Play Extendable Architecture in Figure 3): Users can extensively tailor data center designs, adjusting elements from server specifics to HVAC details via a JSON object file. Moreover, parameters such as workload profiles and weather data can be modified. These allow the users to rapidly prototype different data center designs, monitor potential temperature hotspots. It also accommodates variations in geography and sustainability metrics.

**Interface** (Open AI Gym Controller Interface in Figure 3) To facilitate single as well as holistic multiagent optimization of Data Center Carbon footprint, DCRL-Green provides interfaces for applying reinforcement learning-based control (Control Agents in Figure 3). It uses the massively scalable RLLib (4) that allows users to readily try an array of single as well as multiagent algorithms for optimizing flexible load scheduling, HVAC setpoint control, and battery supply (23; 21; 15; 22; 24).

Overall, DCRL-Green fosters sustainable data center operations, promoting collaborative green computing research within the ML community.

## Applications of DCRL
### Single and Multi-Agent Optimization

DCRL has also been used to optimize HVAC setpoint control using classical single-agent reinforcement learning Table 1 leading to 7% carbon emission reduction and multiagent reinforcement learning Table 2 leading to 13% carbon emission reduction.

| Controller | Energy Footprint (MWh) | Carbon Footprint (TonnesCO2) |
|---|---|---|
| PPO | **26.42** | **13726.56** |
| A2C | 26.74 | 13878.30 |
| ASHRAE Guideline36 | 28.52 | 14839.00 |

Table 1: Energy and Carbon Footprint Comparison for single agent (HVAC Control setpoint)

| Region | Energy | Carbon Footprint |
|---|---|---|
| Arizona | 13.4 ± 0.48 | 13.38 ± 0.62 |
| New York | 13.01 ± 0.12 | 12.77 ± 0.11 |
| Washington | 13.27 ± 0.06 | 13.51 ± 0.05 |

Table 2: Percentage Reduction in Energy and Carbon footprint using a Multiagent approach (MADDPG (5)) when compared against the ASHRAE G36 control

## Conclusion

The paper presents DCRL-Green, an OpenAI Gym environment for reinforcement learning in data centers (DCs), offering customizable DC configurations and targeting various sustainability goals, thereby enabling ML researchers to mitigate climate change effects from rising DC workloads. As part of future work, CFD neural surrogates can automate parameter generation for a custom data center configuration. Lowering the energy usage of data centers and shifting the energy consumption to periods when more green energy is available on the power grid can significantly impact the reduction of the carbon footprint of the data centers and help fight climate change. Also this open source implementation democratizes access to machine learning researchers to contribute to a greener planet.